\documentstyle[11pt,newpasp,twoside,epsf]{article}

\def\edcomment#1{\iffalse\marginpar{\raggedright\sl#1\/}\else\relax\fi}
\reversemarginpar

\def\ms{M$_{\odot}$}
\begin{document}
\title{Evolution of CNO abundances in the Universe}
 \author{Nikos Prantzos}
\affil{Institut d'Astrophysique de Paris, 98bis Bd Arago, 75014 Paris, France}

\begin{abstract}
After summarizing the most important features of current stellar yields
of CNO elements (including recent results concerning rotating and mass losing stars)
I discuss how these yields may help to interpret relevant
observations in the local Galaxy, the Milky Way disk and extragalactic systems
(extragalactic HII regions and DLAs).

\end{abstract}

\section{Introduction}

The study of the production and evolution of the most abundant metals in the Universe, namely
the CNO elements, has profound implications for our understanding of nucleosynthesis and mixing
in the whole stellar mass range, and of the evolutionary status of the galactic systems
where these elements are observed.

In this short review I describe the most important features of current stellar yields
of CNO elements\footnote{My choice of sets of stellar yields is by no means exhaustive: it
is focused on studies where metallicity effects are explicitly taken into account, as well
as on recent results concerning rotating stars;  I apologise
for not discussing in detail other sets of yields that appeared in the literature.}
(Sec. 2) and I discuss how these yields may help to interpret relevant
observations in the local Galaxy (Sec. 3), the Milky Way disk (Sec. 4) and extragalactic systems
(Sec. 5).

\begin{figure}[t!]
\plotfiddle{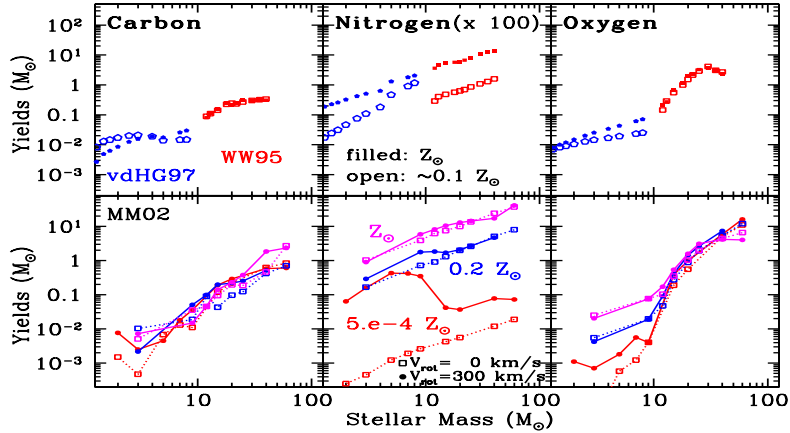}{5cm}{-90}{40}{30}{-150}{175}
\caption{ Stellar yields (ejected mass, in M$_{\odot}$) of C12 ({\it left}), 
N14 ({\it middle}) and O16 ({\it right}), as a function
of stellar mass and metallicity. {\it Top:} Yields of van den Hoek and Grownewegen 
(1997, vdHG97) for low and intermediate mass stars and from Woosley and Weaver 
(1995, WW95) for massive stars; {\it filled symbols} correspond to metallicity 
Z$_{\odot}$, while {\it open symbols} to 0.05 Z$_{\odot}$
for vdHG97 and 0.1 Z$_{\odot}$ for WW95.
{\it Bottom:} Yields of Meynet and Maeder (2002, MM02) 
for 3 different metallicities (Z$_{\odot}$,
0.2 Z$_{\odot}$ and 5 10$^{-4}$ Z$_{\odot}$), clearly identified in the case 
of N (middle panel);
they are given for non-rotating stars ({\it open symbols} connected 
by {\it dotted curves}) and
for stars rotating at 300 km/s ({\it filled symbols} connected by 
{\it solid curves}).}
\end{figure}

\section{CNO yields}

Several recent studies contributed to a substantial improvement of our 
understanding of stellar nucleosynthesis,
in both massive and intermediate mass stars (e.g.
Woosley and Weaver 1995, Thielemann et al. 1996, Limongi et al. 2000, 
Marigo 2001, van den Hoek and Grownewegen 1997,  Meynet and Maeder 2002).
However, despite continuous  refinement in
the input physics of the stellar models, 
important uncertainties still remain, concerning the nuclear
physics, the various mixing processes and the mass loss, especially
during the advanced evolutionary phases. 
As far as CNO yields are concerned, the most important nuclear uncertainty
stems from the $^{12}$C($\alpha,\gamma$) reaction rate 
(e.g. Arnould, these proceedings).
On the other hand, the amount of the mixing depends on the adopted
instability criteria (e.g. Arnett, this volume), as well as on
the treatment of rotation
(e.g. Meynet, these proceedings).

Among  the CNO elements, oxygen is apparently
less affected by these uncertainties;
its yield varies by less than 50\% between the various recently published 
studies (Woosley and Weaver 1995, Thielemann et al. 1996, Limongi et al. 2000).
Carbon is considerably more affected: in the most massive stars 
(above $\sim$30 \ms) the amount of mass
loss becomes quite important at high  metallicities and 
may considerably modify the carbon yields (Maeder 1992);
in intermediate mass stars, ``hot-bottom'' burning (HBB), 
which depends (in a presently poorly understood manner) on
stellar mass and metallicity, constitutes the most important factor of 
uncertainty. In both cases, however, carbon is produced as {\it primary}, i.e.
it is created from the initial hydrogen+helium content of the stars, and
its yield does not vary by orders of magnitude as a function of metallicity.

Nitrogen is a different story. It suffers from the same uncertainties
as carbon, which affect not only its yield but also its very nature as
primary or secondary. Although {\it secondary in principle} (being produced by the
initial C and O of the stars, through the CNO cycle), 
it may be produced also as {\it primary} 
whenever carbon produced by He-burning inside the star
is mixed in hydrogen-rich zones, where the CNO cycle operates. 
This may happen in the case of HBB in intermediate mass stars, as well as in 
rotating stars of all masses according to the recent results of 
Meynet and Maeder (2002, MM02).

Fig. 1 displays yields from recent stellar  nucleosynthesis calculations,
covering the whole stellar mass range and an extensive range in metallicities.
It can be seen that: in intermediate mass stars, HBB (vdHG97) produces
almost primary nitrogen (``almost'', because the yields are not really
independent of metallicity); in non-rotating stars of all masses, nitrogen
is always produced as secondary {\it if} neither HBB nor rotation are 
included (MM02); finally,
in rotating stars of all masses, {\it primary} 
nitrogen is produced 
through rotational mixing (MM02), the effect being more pronounced in low 
metallicities and for intermediate mass stars.\footnote{The MM02 calculations 
do not reach the AGB phase and thus do not include any contribution from HBB burning.}.

The metallicity dependence of carbon yields from the most massive stars
(due to mass loss) is also apparent in Fig. 1. However, this dependence is 
less pronounced than in the case of the Maeder (1992, M92) yields, which concerned 
non-rotating stars with higher mass loss rates than those adopted in MM02.
The M92 yields from massive stars
provided a quite satisfactory fit to the observed evolution of the C/O 
abundance ratio, as shown by  Prantzos et al. (1994).
However, the new yields of MM02 supersede the old ones of M92 and
we shall adopt them in the following.

\section{Evolution of CNO in the solar neighborhood}

From an inspection of the observational data on CNO abundances in halo and disk
stars in the solar neighborhood (Fig. 2) it appears that:

(1) O/Fe is $\sim$ 3 times solar during the halo phase ([Fe/H]$<$-1) and declines
smoothly during the disk phase ([Fe/H]$>$-1); since the O yield is metallicity
independent, this means that some late source of Fe produces
$\sim$2/3 of solar iron. Since the timescale for halo formation is evaluated to $\sim$1
Gyr, this sets the timescale for that late source to become important Fe conributor;
simple models for the rate of type Ia supernovae (SNIa) suggest that these objects
can indeed be the late Fe source. \footnote{It should be noted, however, that the rate of
SNIa is difficult to calculate from first principles, since the very nature of those
systems (i.e. progenitors, accretion rates and timescales etc.) 
is poorly understood at present. The observed decline of O/Fe in the solar
neighborhood provides a useful local constrain, but does not {\it prove} the
correctness of any formula evaluating the SNIa rate.}

(2) C/Fe is always solar (albeit with a considerably more important scatter than O/Fe).
Combined to point (1), this means that a late source of carbon is required during the disk
phase, in order to match the late source of Fe; approximately 2/3 of solar C should be produced
by that late source.

(3) The behaviour of N/Fe at low metallicities is not clear at present. 
Carbon et al. (1987) found that the N/Fe declines below its otherwise solar value for
metallicities [Fe/H]$<$-2; however the authors suggest that, when corrected for poorly
understood effects of T$_{eff}$, that ratio remains approximately 
solar even down to the lowest metallicities. In any case,
a late source of N is required to match the late Fe source in the disk and keep N/Fe
always to its solar value.

(4) Combining (1), (2) and (3) one sees that C/O and N/O should increase by a 
factor of $\sim$2-3 during the disk phase; whatever the late sources of C and N, they 
must produce at least twice as much C and N relatively to O as stars at low metallicities.

\begin{figure}[t]
\plotfiddle{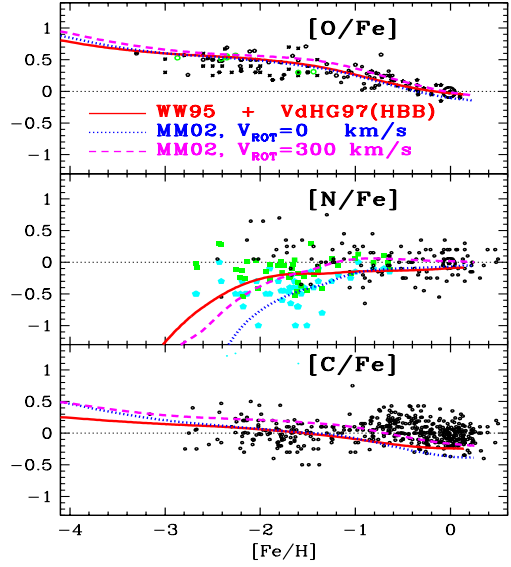}{7cm}{0}{35}{30}{-120}{-15}
\caption{Evolution of [X/Fe] abundance ratio, where X stands for O ({\it top}),
N ({\it middle}) and C ({\it bottom}), as a function of [Fe/H] in the Milky Way (halo
and local disk). The three model curves correspond to yields from vdHG97+WW95 ({\it solid}),
MM02 for non-rotating stars ({\it dotted}) and MM02 for rotating stars ({\it dashed});
see Fig. 1 for yields.}
\end{figure}

In Fig. 2 observations are compared to the results of simple models for the
chemical evolution of the solar neighborhood. These models fullfill all the major local
observational constraints (age-metallicity relation, metallicity distributions of halo and 
disk stars, gas fraction etc.), as explained in detail in Goswami and Prantzos (2000).
Three different sets of stellar yields are adopted: (a) those of vdHG97 for intermediate
mass stars and from WW95 for massive stars, (b) those of MM02 for non-rotating stars
in the whole stellar mass range and (c) those of MM02 for rotating stars (see Fig. 1).

It should be noted that in all three cases the evolution of X/Fe abundance ratio
(where X stands for C, N and O) is not calculated in a self-consistent way.
For the first set of yields the reason is that different  input physics have been
used in the calculations of vdH97 and WW95 (the most important being the
$^{12}$C($\alpha,\gamma$) rate). On the other hand, MM02 calculate the whole stellar
mass range with the same physics, but they do not go beyond carbon burning in
massive stars, and thus they cannot provide yields for Fe; one has then to make 
assumptions about the corresponding Fe yields and here we adopted those of WW95 
as a function of stellar mass and metallicity (interpolating them in the
corresponding grid values of MM02). Clearly, there is an inconcistency in the
yields, since mass loss (not included in WW95 calculations) may affect the size of
the Fe core, the mechanism of the explosion and the final Fe yield. However, 
taking into account all the uncertainties associated with the explosion of Type II
supernovae and the subsequent fall-back (see e.g. WW95), we feel that our treatment
in cases (b) and (c) does not introduce more uncertainties than those inherent in the
WW95 yields.

\begin{figure}[t]
\plottwo{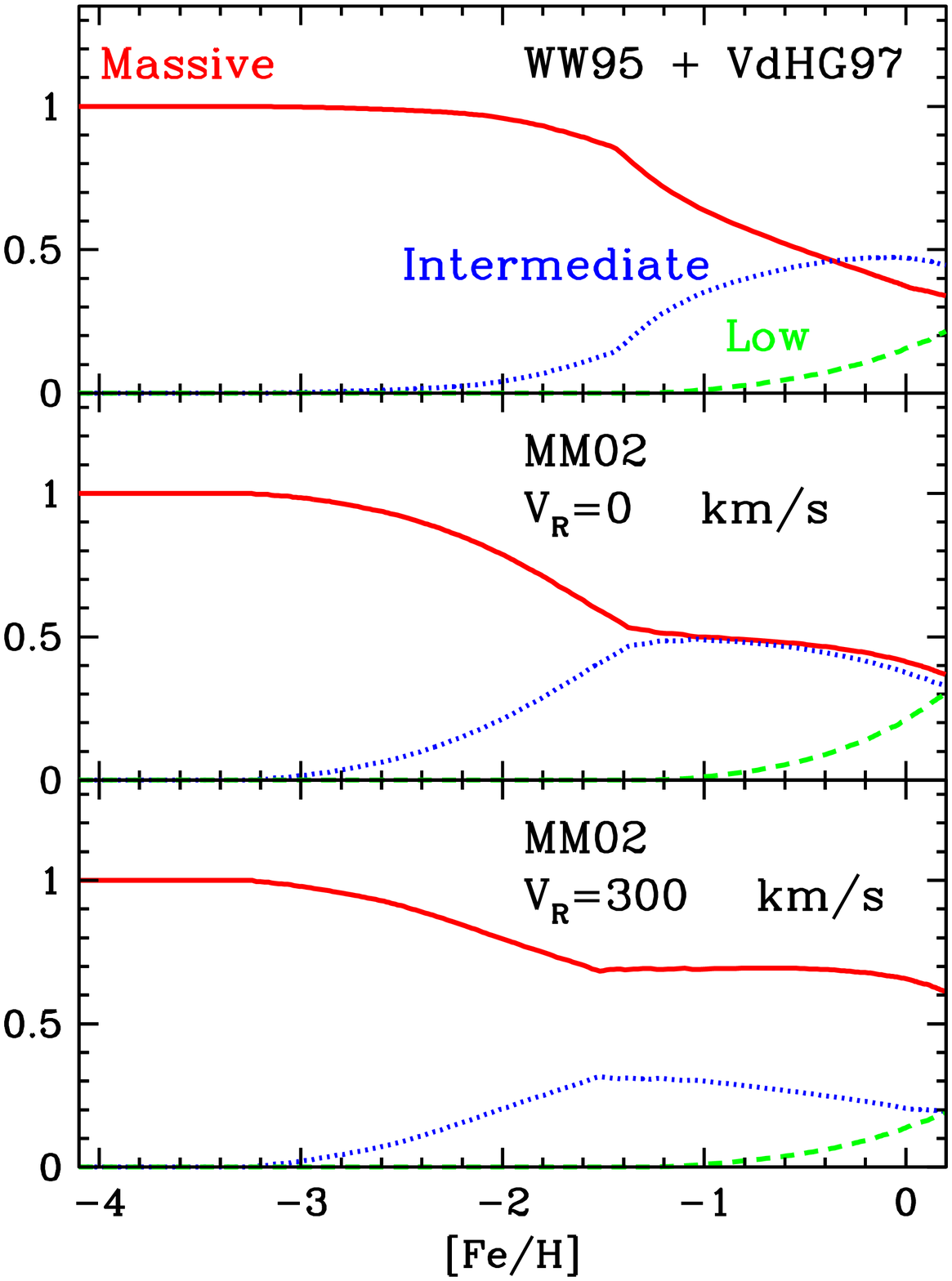}{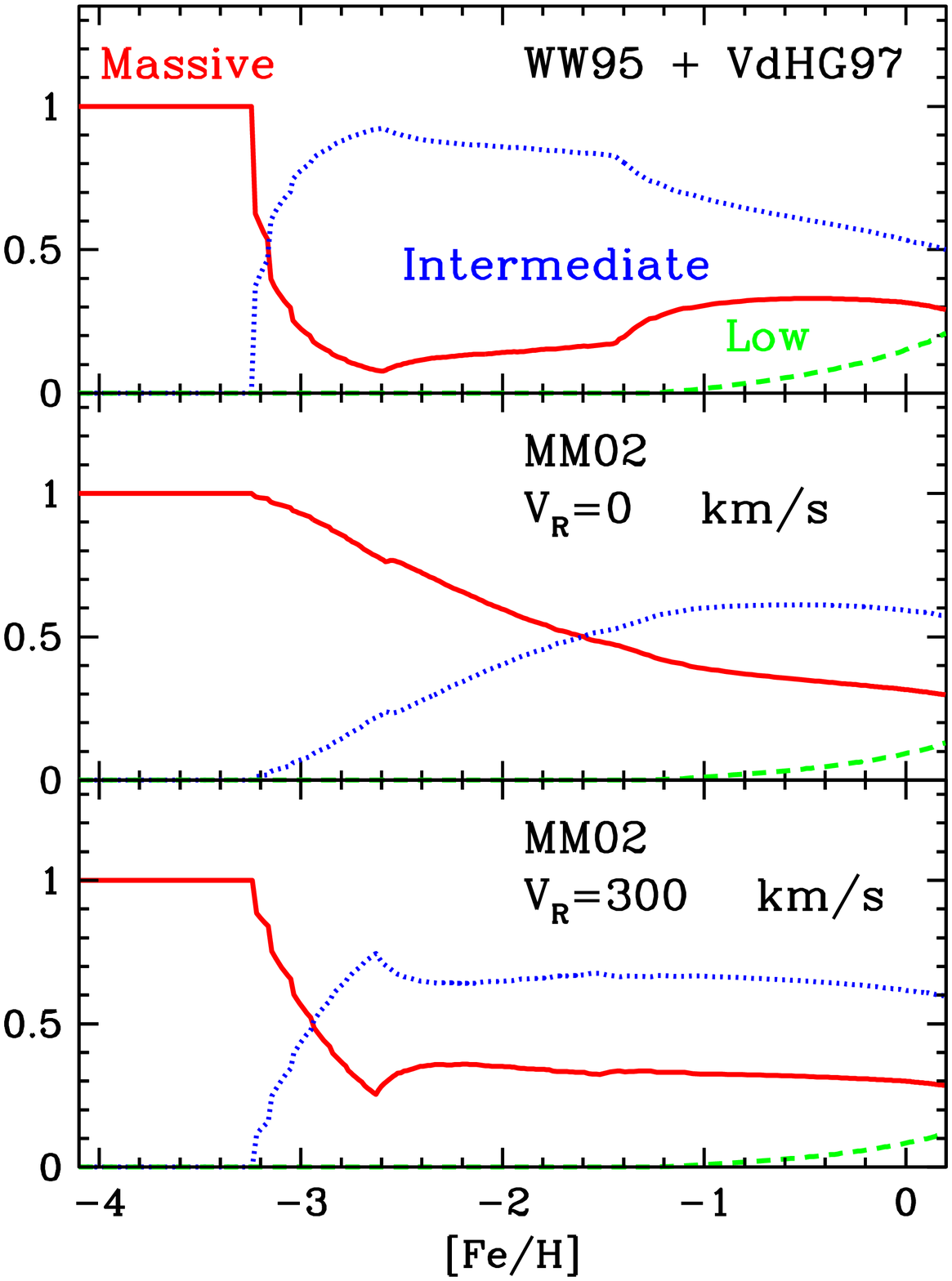}
\caption{Contribution of various stellar mass ranges to the local galactic production
of C ({\it left}) and N ({\it right}) as a function of [Fe/H].
The three panels display results obtained with yields from vdHG97+WW95 ({\it top}),
MM02 for non-rotating stars ({\it middle}) and MM02 for rotating stars ({\it bottom}).
In all panels the contributions of massive stars (M$>$10 M$_{\odot}$), intermediate
mass stars (2$<$M/M$_{\odot}<$9) and low mass stars (M$<$2 M$_{\odot}$) are indicated
by {\it solid}, {\it dotted} and {\it dashed} curves, respectively.}
\end{figure}

From Fig. 2 it can be seen that:

(1) The three sets of O yields from massive stars (WW95 with no mass loss, MM02 with mass loss
and with or without rotation) lead to quasi-identical  results.

(2) The three sets of yields lead to slightly different results for carbon, but well within
the scatter of presently available observations. In all three cases, late production of C 
{\it almost} matches late Fe production by SNIa ({\it almost}, because at solar birth C/Fe
is subsolar in all three cases).
The late source of C is IMS for sets (a) and (b) of adopted yields, and massive rotating -
and mass losing - stars for set (c). This is clearly illustrated in Fig. 3 ({\it left}) 
where the contributions of various
stellar mass ranges to C production is displayed. Note that the M92 yields provided a better
fit to the observed evolution of C (Prantzos et al. 1994, Gustaffson et al. 1999) but they
are superseded by the MM02 yields.

(3) In all three cases nitrogen evolves as secondary in the very early Galaxy, up to [Fe/H]=-2
for yield sets (a) and (c) and up to [Fe/H]=-1 for set (b). In cases (a) and (c) there is
quasi-primary N production (from HBB and from rotational mixing, respectively), which flattens
the N/Fe ratio in the -2$<$[Fe/H]$<$-1 range. Finally, for [Fe/H]$>$-1, low mass
IMS (2-3 \ms) dominate N production 
and release quasi-secondary N (since the yield dependence on metallicity is stronger at
high metallicities), which matches the late Fe production from SNIa. The contribution of each
range of stellar masses can be seen in Fig. 3 ({\it right}).

\begin{figure}[t]
\plottwo{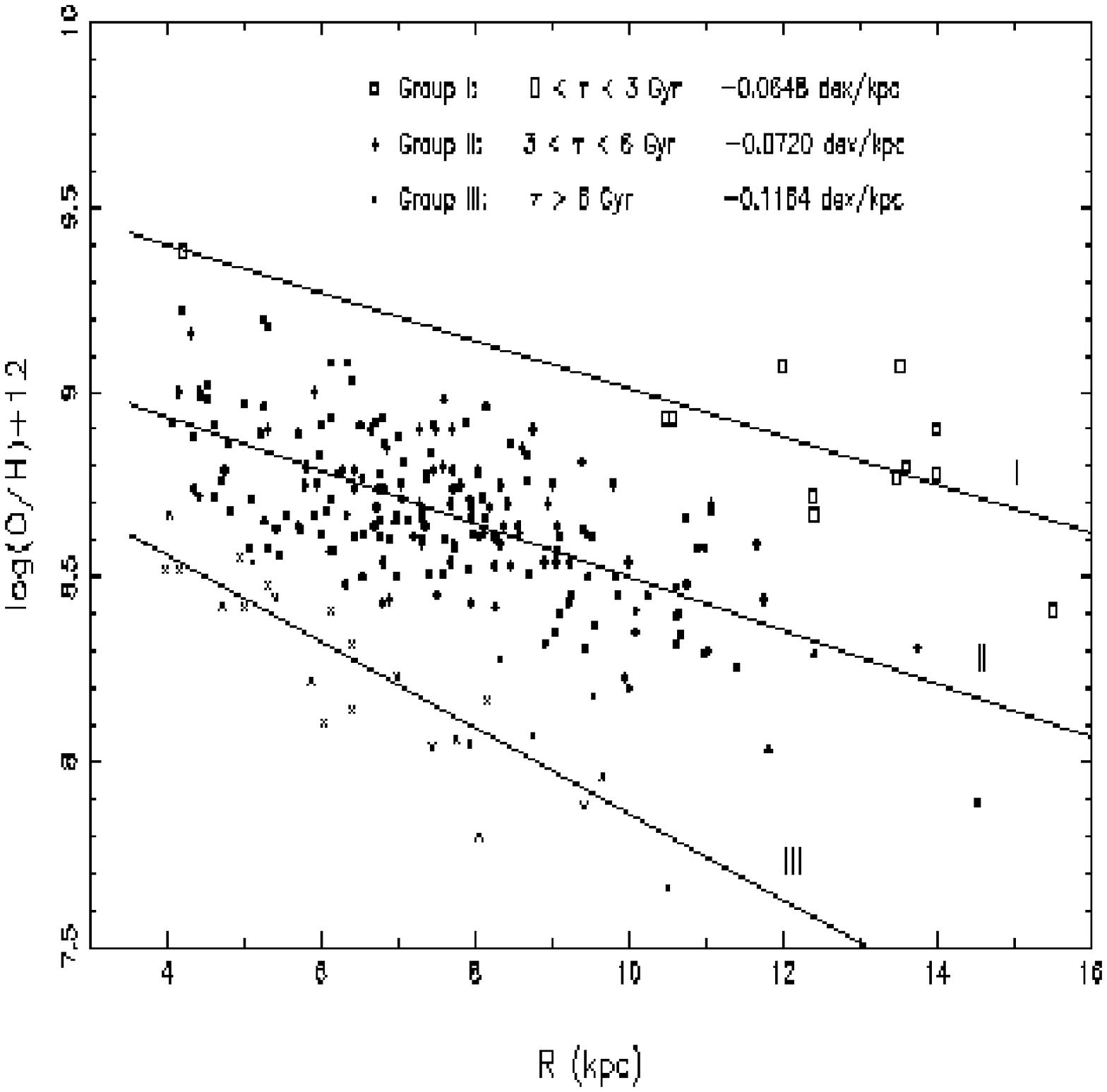}{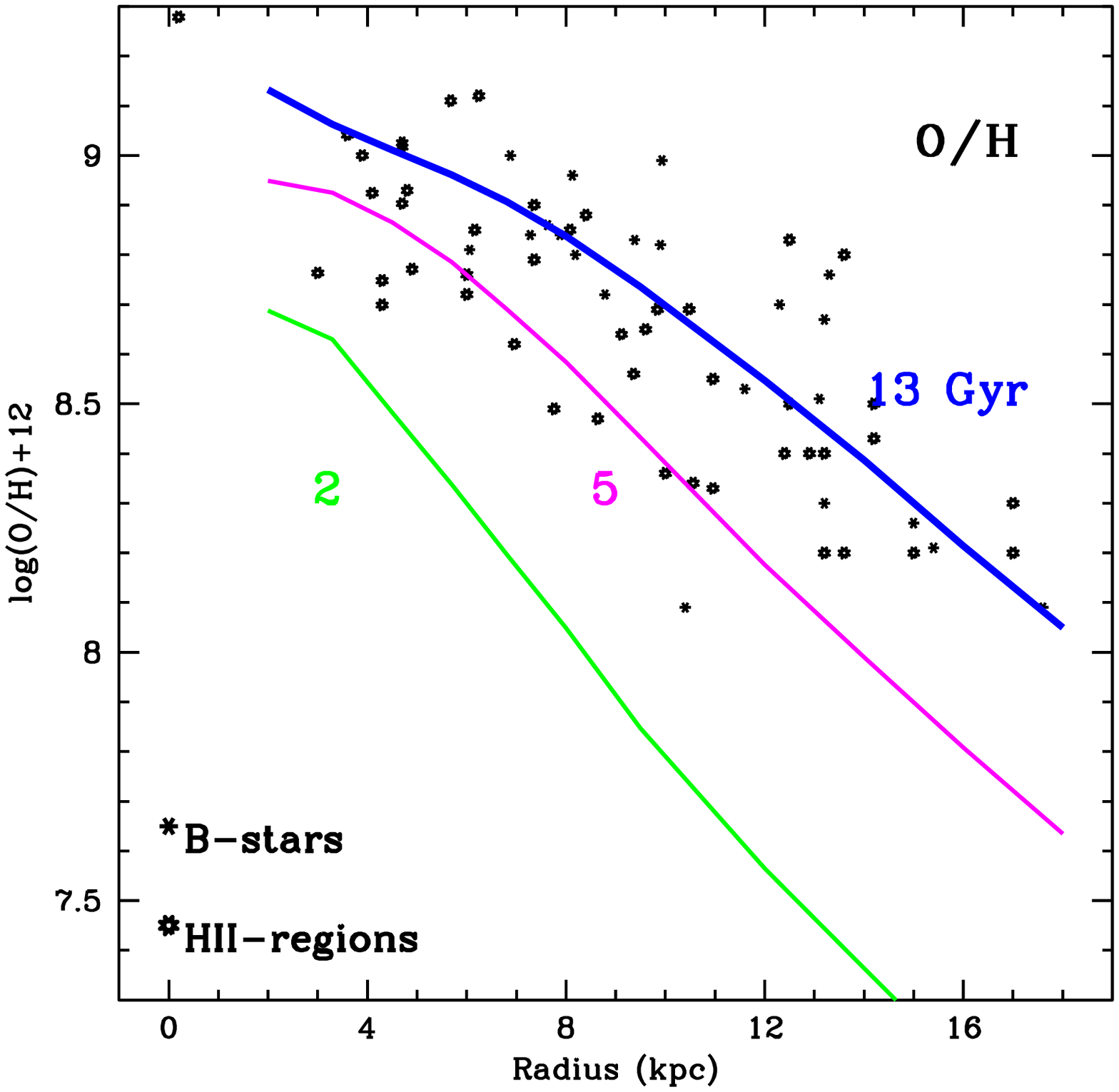}
\caption{{\it Left:} Observed abundance gradients of O/H in the Milky Way disk, traced by
planetary nebulae of various classes (I, II and III) and corresponding ages
(from Maciel et al. 2002). {\it Right:} Model abundance gradients of O/H in the Milky Way disk
at 2, 5 and 13 Gyr, respectively({\it curves})  
and comparison of the latter to observed
present-day abundance gradient, traced by HII regions and B-stars ({\it data points},
from Hou et al. 2000).}
\end{figure}

The conclusions of this section can be summarized as  follows:

i) The N yields of rotating stars of MM02 lead to similar results as those of vdHG97 with
HBB.

ii) IMS always dominate N production in the Milky Way disk. For models with HBB or rotational mixing
they also dominate down to [Fe/H]=-3.

iii) In the framework of simple models, there is no way to obtain solar N/Fe at [Fe/H]=-3;
the secondary production of N from massive stars dominates at those early times.
Current stellar yields and ``standard'' models of galactic chemical evolution
match the non-corrected data of Carbon et al. (1988), as also found
in Liang et al. (2001) or Chiappini et al. (2002). 
However, if the corrected data of Carbon et al. (1988)
represent ``reality'', then either:

- (i) a mecanism should be found for substantial primary
N production in massive stars,  or 

- (ii) the timescales obtained in simple GCE evolution models
should be revised, allowing for IMS (and their quasi-primary N)
to enter the galactic scene even before [Fe/H]=-3 (see Prantzos 2003 for such a revision).

\begin{figure}[t]
\plotfiddle{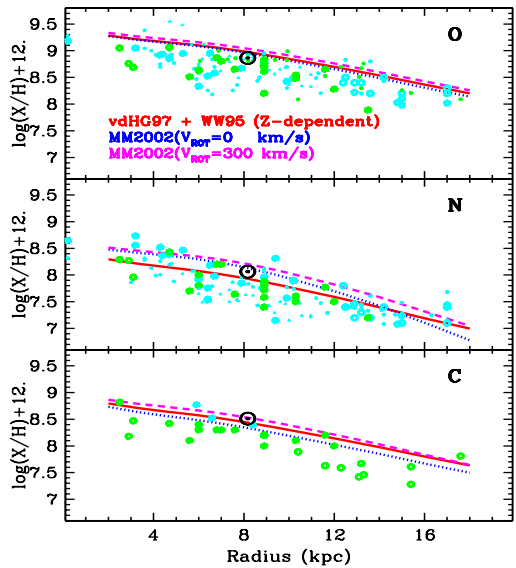}{7cm}{0}{35}{30}{-120}{-20}
\caption{Present day abundance gradients of C ({\it bottom}), N ({\it middle}) and
O ({\it top}) in the Milky Way disk. Observational data correspond to abundances
of HII regions and B-stars. Model results are obtained with yields from vdHG97+WW95 ({\it solid
curves}), non-rotating stars of MM02 ({\it dotted curves}) and rotating stars of MM02 
({\it dashed curves}).}
\end{figure}

\section{Evolution of CNO in the Milky Way disk}

Maciel et al. (2002) provided recently evidence that the oxygen abundance gradient 
in the Milky Way disk was steeper
in the past, by measuring it in planetary nebulae (PN) of various morphological types and age classes
(Fig. 4, {\it left}).  This result may have important implications for our understanding of 
the formation of the Galactic disk,  but three points should be made first: 
(1) the uncertainties in evaluating ages of planetary nebulae are quite substantial, 
(2) the absolute oxygen values of ``group I'' PN in Maciel et al. (2002) are a factor of $\sim$2
higher  than values of other young disk objects (e.g. B-stars and HII regions)
and (3) the present-day value of the oxygen abundance gradient in the Milky Way is still subject to
considerable debate: the''canonical'' value of d[O/H]/dR = -0.07 dex/kpc (see Hou et al. 2000
and references therein) could be as low as -0.04 dex/kpc (see Deharveng et al. 2000, 
Cunha et al. these proceedings). 

Clearly, the issue is far from being settled observationally
yet, but an important first step has already been made: at least qualitatively, the
observed evolution of the abundance gradient is in agreement with models in which
the disk is formed inside-out (e.g. Molla et al. 1997, Boissier and Prantzos 1999,
Allen et al. 1998, Hou et al. 2000). The results of such a model (satisfying all
the major observational constraints of the Milky Way disk) are displayed in
Fig. 4 ({\it right}) at three different ages (2, 5 and 13 Gyr, respectively).
The latter agrees well with the ``canonical'' value of -0.07 dex/kpc, inferred from
observations of HII regions and B-stars.

Assuming that the evolution of the oxygen abundance gradient is well understood,
we display in Fig. 5 the present-day abundance gradients of CNO elements,
obtained with the same sets of yields (a, b and c) as in Sec. 3 and the model
of Hou et al. (2000) for the Milky Way disk evolution. 
It can be seen that:

(i) the different sets of yields leads to similar results in the cases
of O and C and marginally different ones in the case of N; in the latter case, only yield
set (b) (with purely secondary N) leads to a substantially different
(steeper) abundance gradient, but this yield set is implausible in view
of the results discussed in Sec. 3.

(ii) observational scatter at all radii is much more important than differences
produced by the different sets of yields; this situation also holds for abundance
ratios (C/O or N/O vs galactocentric radius) as shown in Hou et al. (2000).
 
In summary, current data of CNO abundances across the Milky Way disk
can be explained with current yields and ``reasonable'' simple chemical 
evolution models, but they do not offer much useful insight on 
the yields (in particular, about any metallicity dependence of the yields
at higher than solar metallicities, such as those prevailing in the inner disk); 
much more observational work on 
the abundance ratios, especially  in the inner disk, is required for that.
Finally, note that  the adopted sets of yields do not extend to metallicities higher
than Z$_{\odot}$, making the modelisation of N evolution in the inner disk inaccurate.

\section{CNO in extragalactic HII regions and DLAs}


Abundances of C, N and O have been observed through emission lines in 
extragalactic HII regions and starbursts
in  the local Universe (e.g. Izotov and Thuan 1999, Pilyugin et al. 2002, Mouhcine and Contini 2002) 
and through absorption lines in remote clouds of neutral hydrogen
(DLAs, Prochaska et al. 2002, Pettini et al. 2002). 
Note that our current understanding of the nature of the
corresponding galactic systems is poor in the former case and less than poor in
the latter. In those conditions, any attempt to constrain the properties
of those systems by CNO observations {\it alone} 
seems rather futile, at least until the
intricacies of CNO nucleosynthesis are well understood (i.e. evolutionary
timescales and appropriate yields of the sites of primary nitrogen and of 
late carbon production).

\begin{figure}[t]
\plottwo{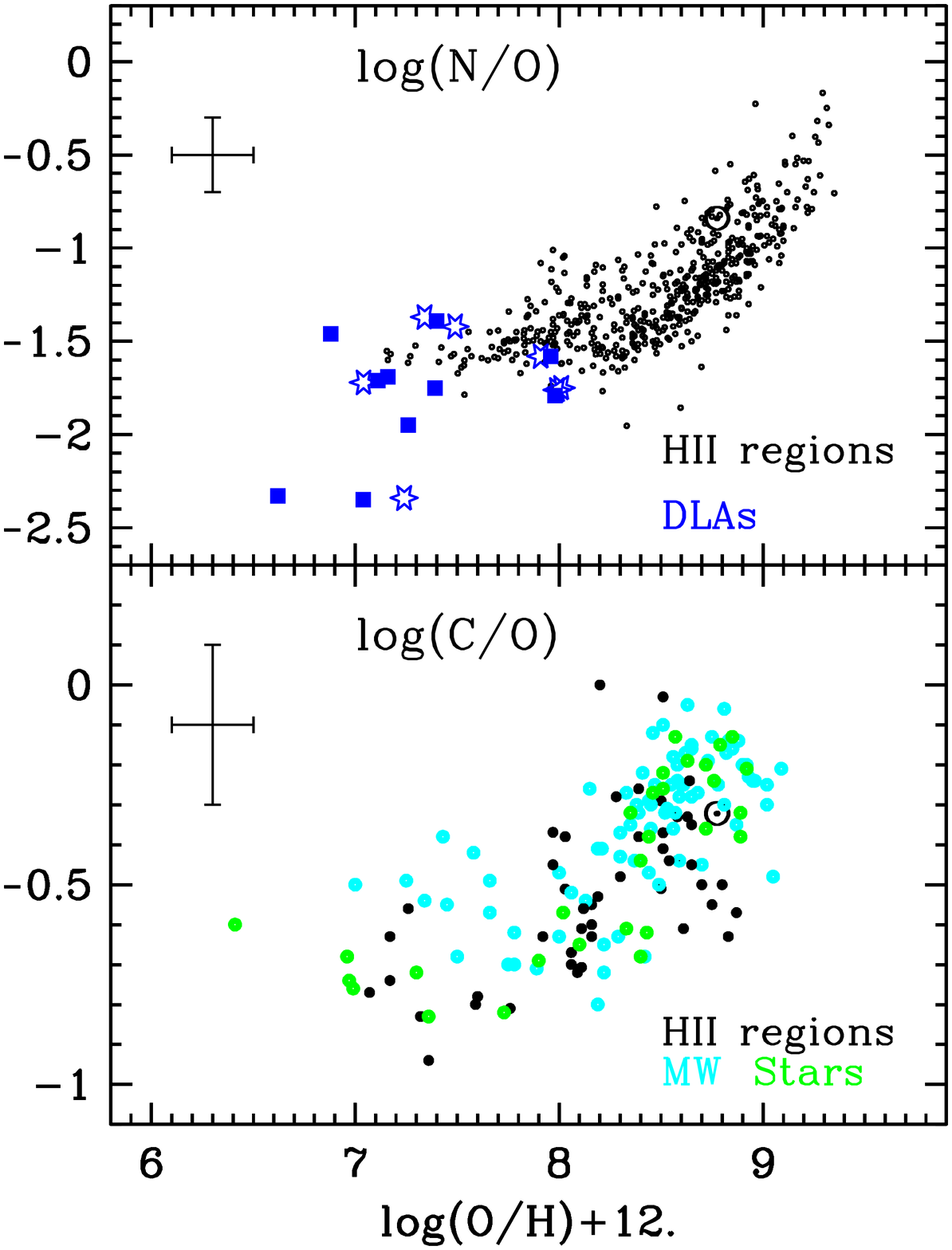}{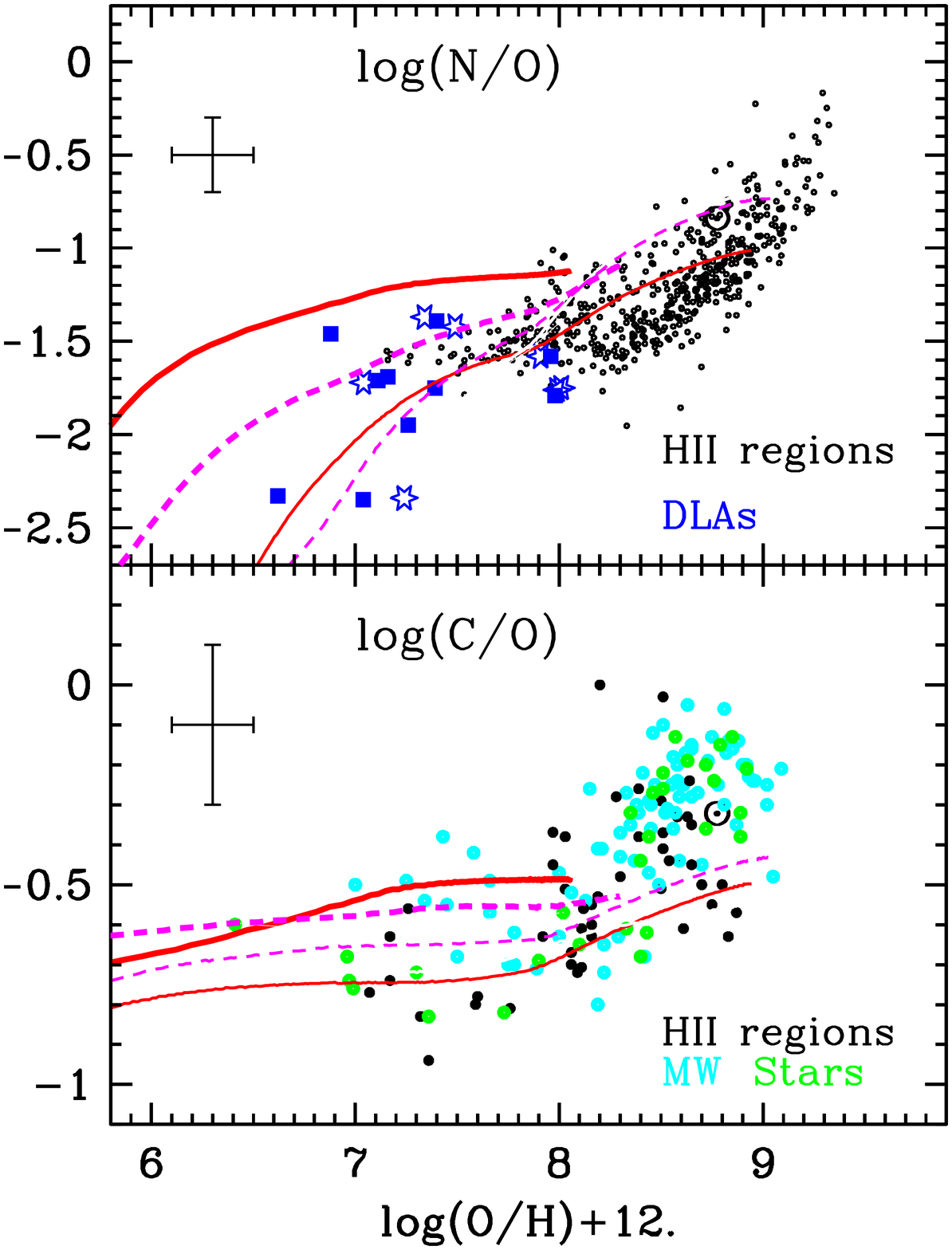}
\caption{{\it Right:} Observations of N/O ({\it top}: {\it small symbols} for HII regions
and {\it large symbols} for DLAs) and C/O ({\it bottom}: {\it dark symbols} for HII regions
and {\it light symbols} for MW stars) abundance ratios
in various objects as a function of O/H. 
{\it Left:} Comparison to 2 simple (and probably irrelevant!) models: 
i) a solar neighborhood model ({\it  thin curves}, same as in Fig. 2) and 
ii) the same model, but 
with the star formation efficiency reduced by  a factor of ten ({\it thick curves}), reaching
lower metallicities. {\it Solid curves} correspond to the vdHG97+WW95 yields
and {\it dashed curves} to the MM02 yields of rotating stars.}
\end{figure}

The relevant observations are presented in Fig. 6 ({\it left}) for N/O and C/O, both as
a function of O/H. In the case of N/O, data for extragalactic HII regions 
({\it small symbols}) reveal 
an increase at high O/H (approximately above log(O/H)=-4) and a ``plateau''
at lower O/H values, trends that correspond to a ``secondary'' and a ``primary'' nitrogen
production, respectively; in both cases, the lifetimes (and masses) of the corresponding
nucleosynthesis sites are unknown. The small scatter of N/O along the ``plateau'', obtained mainly 
by Izotov and Thuan (1999; see also Izotov, these proceedings) is intriguing, especially 
when contrasted with the large scatter obtained at higher O/H values. Some of the DLA
data ({\it large symbols}) 
fall also along the plateau of the HII-region values, while others appear clearly
below the plateau, by about 0.6 dex on average; this gave rise to arguments about a ``bimodality''
of the N/O values in DLAs (e.g. Prochaska et al. 2002; also Henry, these proceedings), although
the statistics of 
the presently available data are clearly insufficient for such a conclusion
(see also Molaro, these proceedings).

Similar, albeit not identical features, are observed in the case of C/O. Both HII regions
({\it dark symbols}) and MW stars ({\it light symbols}) show an increase of C/O above O/H$\sim$-4
and a ``plateau'' below that value; however, the rise is smaller than in the case of N/O
(only a factor of $\sim$3, compared to a factor of $\sim$10) and suggests only 
a supplementary late source of C (as argued in Sec. 3) but clearly 
not a secondary behaviour for that element. Also, unlike the case of N/O, there
is a large scatter around the plateau values, comparable to the one at higher O/H.

On the right part of Fig. 6 we check whether these data can be interpreted in the
framework of simple-minded models of GCE \footnote{This is {\it not}
the same thing as to present {\it models} for the corresponding galactic systems; as stressed in
the beginning of Sec. 5, it is impossible to model a galactic system based {\it only} on CNO
abundance data.}. The model adopted for the local evolution of the MW (also shown in Fig. 2)
is displayed with {\it thin curves} for yield sets (a) and (c); the model reaches solar
O/H values, N/O behaves mostly as secondary (this is {\it not} obvious when N/Fe is plotted
vs. [Fe/H], since Fe production by SNIa largely matches the secondary N production) and
C/O increases slightly at high O/H\footnote{The M92  yields of carbon
matched the data perfectly, while the Padova yields - not discussed here - are shown to match the data
well in Carigi (2002).}.
{\it Thick curves} present the same model, with the star formation efficiency reduced
by a factor of ten: only low values of O/H are reached and consequently the
secondary N component (dominant at high metallicities) does not show up; 
quasi-primary N from IMS dominates.

The results presented in  Fig. 6 suggest  that the high (``plateau'') values of N/O at low O/H
should be attributed to systems old enough for IMS to contribute and  with low
SF efficiencies. The low N/O values of DLAs should be attributed to relatively ``young'' systems,
polluted only by the secondary N of massive stars (see also Pettini et al. 2002). Further observations
are required to decide whether the current ``gap'' between low and high N/O values in DLAs is
significant or not.

\medskip
\acknowledgements{{\small I am grateful to the organisers of the CNO Workshop for invitation 
and financial support, and 
to Dick Henry for providing CNO data and for enlightening discussions
during the workshop.
I am indebted to Andr\'e Maeder, for constantly providing new insights into the workings of stars
and their possible impact on galactic chemical evolution. {\it Bonne anniversaire Andr\'e !}}
 {\small
{}
}
\end{document}